\documentstyle[12pt,epsf]{article}
\newcommand{\beq}{\begin{equation}}
\newcommand{\eeq}{\end{equation}}
\newcommand{\beqa}{\begin{eqnarray}}
\newcommand{\eeqa}{\end{eqnarray}}

\begin{document}

\noindent REVISED VERSION \hfill KFA-IKP(TH)-1997-02 

\hfill TK 97 02

\hfill hep-ph/9701260

\bigskip\bigskip\bigskip

\begin{center}

{{\Large\bf Baryon magnetic moments in \\[0.3cm] chiral perturbation
    theory}\footnote{Work supported in part by funds provided by the
    Graduiertenkolleg "Die Erforschung subnuklearer Strukturen der Materie".}}

\end{center}

\vspace{.2in}

\begin{center}
{\large  Ulf-G. Mei{\ss}ner$^\ddagger$\footnote{email: 
Ulf-G.Meissner@fz-juelich.de}, S. Steininger$^\dagger$\footnote{email: 
sven@pythia.itkp.uni-bonn.de}}

\bigskip

\bigskip

$^\ddagger${\it Forschungszentrum J\"ulich, 
Institut f\"ur Kernphysik (Theorie)\\ 
D-52425 J\"ulich, Germany}

\bigskip

$^\dagger${\it Universit\"at Bonn, Institut f{\"u}r Theoretische Kernphysik\\
Nussallee 14-16, D-53115 Bonn, Germany}\\

\end{center}

\vspace{.7in}

\thispagestyle{empty} 

\begin{abstract}
\noindent
We consider the chiral expansion of the octet baryon magnetic moments in
heavy baryon chiral perturbation theory including {\it all} terms which 
are of order $q^4$. These terms are formally of quadratic
order in the quark masses. We show that despite  the large non--analytic 
quark mass corrections to the Coleman--Glashow relations at order $q^3$,
including all analytic and non--analytic  corrections at order $q^4$,
which in total are of moderate size, allows for a  fit 
to the measured magnetic moments due to the appearance of counter terms
with free coupling constants of natural size. In this scheme,
the $\Lambda \Sigma^0$ transition moment is predicted to be
$\mu_{\Lambda \Sigma^0} = (1.42 \pm 0.01)\, \mu_N$. 
\end{abstract}

\vspace{.3in}

\centerline{PACS: 11.30.Rd, 12.39.F, 13.40.Em}

\centerline{Keywords: 
{\it Baryon magnetic moments}, {\it chiral perturbation theory}}

\vfill

\newpage

\section{Introduction}

The magnetic moments of the octet baryons  have been measured with
high precision over the last decade. On the theoretical side, SU(3)
flavor symmetry was used by Coleman and Glashow \cite{CG} to  
predict seven relations between the eight moments of the $p,n,\Lambda,
\Sigma^\pm, \Sigma^0, \Xi^-, \Xi^0$ and the $\mu_{\Lambda \Sigma^0}$ 
transition moment in terms of two parameters.  One of these relations is
in fact a consequence of isospin symmetry alone. In modern language, 
this was a tree level calculation with the lowest order effective
chiral meson--baryon Lagrangian of dimension two. Given the
simplicity of this approach,
these relations work remarkably well. Within chiral perturbation
theory, corrections beyond tree level have been calculated treating the
baryons as relativistic spin--1/2 fields \cite{CP}\cite{GSS}\cite{krause}.
As it turns out, the large kaon and pion loop contributions leading to
non--analytic corrections of the type $m_q^{1/2}$ and $m_q \ln m_q$ (with
$m_q$ a generic symbol for any of the light quark masses $m_{u,d,s}$)
tend to worsen the tree level results. With the advent of heavy baryon
chiral perturbation theory, this problem was reconsidered in \cite{jlms}.
In that paper, one loop corrections with intermediate octet and decuplet
states were considered. It turned out, however, that the graphs with
intermediate spin--3/2 states did not cancel the Goldstone boson loop
contributions as it is the case in other observables. It was argued that
smaller axial couplings are needed to reduce the kaon loops, but no firm
conclusion was reached. In addition, one relation amongst the magnetic
moments free of the axial couplings was derived and found to be in good 
agreement with experiment. Furthermore, the baryon magnetic moments were
also considered in the large--$N_C$ \cite{jmNc} and in a combined 
large--$N_C$ and $m_s$ \cite{lmr} expansion leading to some interesting 
relations. In these calculations, an extended power counting scheme was used
such that graphs with intermediate decuplet states start to
contribute already at order $q^3$. Nevertheless, inspection of fig.2
in \cite{jlms} reveals that also terms of higher order $q^4$ (even in
the extended counting scheme), which in part
lead to the non--analytic corrections of the type $m_q \ln m_q$, were
included in addition to the leading order $q^2$ and $q^3$
graphs.\footnote{As we will show, there are some additional terms at
this order related to dimension two insertions 
with fixed coefficients and scalar symmetry breaking.}
Recently, a reordering of the chiral expansion based on linear
flavour symmetry breaking was proposed to overcome the difficulties with
the large non--analytic corrections\cite{bos}. We will show here that
chiral perturbation theory based on the conventional counting scheme
can indeed describe the magnetic moments with
moderate corrections beyond next--to--leading order ${\cal O}(q^4)$,
i.e. they are typically a factor two to three smaller than the  ${\cal O}(q^3)$
ones. Our analysis indicates that
there is no need to take unphysically small axial couplings or include
the spin--3/2 states explicitely. To substantiate this claim, we
perform a {\it complete} calculation at order $q^4$ based on the
standard power counting, taking into
account  vertices from the dimension two meson--baryon Lagrangian,
which are of the type ($1/m\times$~fixed coefficient),
proportional to the scalar second order symmetry breaking and
also the ones of the form $\sim \partial_\mu \phi \, \partial_\nu
\phi$, with $\phi$ denoting the Goldstone bosons.
The pertinent coupling constants related to the latter terms 
can be estimated by resonance saturation as described below. 
Such contributions  have only been considered partly before. If one
includes the decuplet in the effective field theory and expands the
one--loop graphs with intermediate spin--3/2 states, one recovers such
type of operators (and others which are formally of higher order). We
show that there are also important t--channel vector meson
contributions which previously have not been taken into account. 

The manuscript is organized as follows. The necessary formalism is
detailed in section~2, with all lengthy formulae relegated to the appendix.
We discuss  the chiral corrections at orders $q^2$, $q^3$
and $q^4$, respectively, with particular emphasis on the latter type
of contributions. Section~3 contains the estimation
of some low--energy constants. This is based on the resonance exchange 
saturation hypothesis which works well for scattering processes but
is not yet under control for symmetry breaking terms. Therefore, the 
low--energy constants related to the latter type of operators are left as
fit parameters. Our results are presented in section~4
and the summary and conclusions are given in section~5. 

\section{Formalism}
 
The starting point of the heavy baryon 
CHPT \cite{jm}\cite{bkkm} is an
effective Lagrangian formulated in terms of the asymptotic fields,
here the octet of Goldstone bosons and the ground state baryon octet, 
denoted $B$. It admits a low energy expansion of the form
\beq
{\cal L}_{\rm eff} = {\cal L}_M + {\cal L}_{MB} =
{\cal L}_M^{(2)} +  {\cal L}_M^{(4)} +
{\cal L}_{MB}^{(1)} + {\cal L}_{MB}^{(2)}
+ {\cal L}_{MB}^{(3)} + {\cal L}_{MB}^{(4)} + \ldots \,
\eeq  
where the subscript $'M'$ ($'MB'$) denotes the meson (meson--baryon)
sector and the superscript $'(i)'$ the chiral dimension.
The Goldstone bosons are collected in the familiar SU(3)--valued
field $U = \exp\{i \phi / F_\phi \}$, with $F_\phi$ the octet decay
constant. If not otherwise specified, we set $F_\phi = F_K = 1.21\, F_\pi$.
The baryons are  given by a $3\times3$ matrix, transforming 
under SU(3)$_L \times$ SU(3)$_R$ as 
usual matter fields,  $B \to B' = KBK^\dagger$, with $K(U,L,R)$ the 
compensator field which is an element of the conserved subgroup
SU(3)$_V$. We use the notation of \cite{bkmz}.  The ellipsis
stands for terms  not needed here. Note that the terms from ${\cal
  L}_{MB}^{(3)}$ are only needed for the wave-function renormalization.
Beyond leading
order, the effective Lagrangian contains parameters not fixed by
chiral symmetry, the so--called low--energy constants (LECs). In principle,
these LECs should be pinned down from data or calculated by means of
lattice gauge theory. We will resort to the resonance saturation
method to estimate some of these as explained in more detail below.
In what follows, we will work to order $q^4$ in the chiral
expansion. The pertinent tree and one--loop graphs are shown and
catalogued in fig.~1. Consequently, the magnetic moment of any baryon
takes the form
\beq 
\label{mubform}
\mu_B = \mu_B^{(2)} + \mu_B^{(3)} + \mu_B^{(4)} \quad ,
\eeq
where the superscript '$(n)$' denotes the chiral dimension. Let us now
dissect the various contributions.

\subsection{Magnetic couplings at leading order $q^2$}

The magnetic couplings of the photon to the baryons start at chiral
dimension two and the corresponding contribution to the magnetic
moments are calculated from the tree graph shown in fig.~1a.
These are the terms from ${\cal L}_{MB}^{(2)}$ underlying the
Coleman--Glashow analysis
\begin{equation} \label{LMB2}
{\cal L}_{MB}^{(2)} = 
-\frac{i}{4m} \, b_6^F \, \langle \bar{B} [S^\mu,
S^\nu][ F_{\mu \nu}^+, B] \rangle   
-\frac{i}{4m} \,b_6^D \, \langle \bar{B} [S^\mu,
S^\nu]\{ F_{\mu \nu}^+, B\} \rangle   
\end{equation}
with $S_\mu$ the covariant spin--operator, $F_{\mu \nu}^+ =
-e(u^\dagger Q F_{\mu\nu}u+ uQF_{\mu\nu}u^\dagger )$ and $\langle
\ldots \rangle$ denotes the trace in flavor space. 
Here, $Q={\rm diag}(2,-1,-1)/3$ is the quark charge matrix, $u = \sqrt{U}$
 and $F_{\mu \nu}$ the conventional
photon field strength tensor. Throughout we work in the isospin limit
$m_u = m_d$ and thus neglect the small $\Lambda-\Sigma^0$ mixing.
Notice that we normalize the low--energy constants  $b_6^F$ and
$b_6^D$ such that we directly get the total, not just the anomalous,
magnetic moment for the charged particles under consideration. This
simplifies the calculation of some terms at order $q^4$. The
Clebsch--Gordan coefficients for the various octet states are
collected in appendix~A, i.e. all octet magnetic moments are given in
terms of these two LECs. This leads to the Coleman--Glashow relations
(we show here the form advocated in \cite{jlms})
\beqa
\mu_{\Sigma^+} &=& \mu_p \,\, , \,\, 2 \mu_\Lambda = \mu_n \,\, , \, \,
\mu_{\Sigma^-} + \mu_n = -\mu_p \,\, , \nonumber \\
\mu_{\Sigma^-} &=&\mu_{\Xi^-} \,\, , \,\, \mu_{\Xi^0} = \mu_n \,\, , \, \,
2 \mu_{\Lambda\Sigma^0} = - \sqrt{3}  \mu_n \,\, ,
\eeqa
and the isospin relation
\beq 
\mu_{\Sigma^0} = \frac{1}{2} \biggl( \mu_{\Sigma^+} +\mu_{\Sigma^-}\biggr)
\quad .
\eeq  

\subsection{One--loop contribution at order $q^3$}

The first loop corrections arise at order $q^3$ in 
the chiral counting. They are given entirely in terms of the lowest
order parameters from ${\cal L}_M^{(2)} + {\cal L}_{MB}^{(1)}$. These are
$F_\phi$, the quark masses encoded in the meson masses as well as the
two axial coupling constants $F$ and $D$. We use here $F=1/2$ and $D=3/4$.
To order $q^3$, the one loop calculation leads to corrections of the
type (we just show the result for the proton since the pertinent 
coefficients for the other baryons are given in appendix~A, 
see also \cite{CP}\cite{jlms}),
\begin{equation} \label{mup3}
\mu_p^{(3)} = -\frac{m}{8\pi} \biggl[ (D+F)^2 \frac{M_\pi}{F_\pi^2}
+ \frac{2}{3} (3F^2 + D^2) \frac{M_K}{F_K^2} \biggr] \,\, .
\end{equation}
One easily convinces oneself that these loop corrections are large,
e.g. $\mu_p^{(3)} = -2.21$,\footnote{We give all moments in units
of nuclear magnetons $(\mu_N)$ without further specification.} using
the nucleon mass $m_N = 0.94\,$GeV as the average octet mass. 
It is important to note that
for pion (kaon) loops we use the pion (kaon) decay constant. This is
legitimate since the difference between these two is of order ${\cal
  O}(m_q)$ and thus beyond the accuracy we are calculating.
As discussed below, the $q^3$
calculation does not lead to an improvement of the simple tree calculation
based on eq.(\ref{LMB2}). Such a behaviour is already known from the
analogous SU(2) calculation in ref.\cite{bkkm}, where e.g. the  pion 
loop contribution $\sim M_\pi^3$ is large and cancels 2/3 of the 
leading term, which
is the proton anomalous magnetic moment in the chiral limit. It is
also obvious from eq.(\ref{mup3}) that the pion and kaon loops
contribute with similar magnitude. We remind the reader that
 already Caldi and Pagels
\cite{CP} derived three relations which are valid to this order and 
independent of the values of the axial coupling constants,
\beq
\mu_{\Sigma^+} = -2\mu_{\Lambda} -\mu_{\Sigma^-} \,\, , \,\,
\mu_{\Xi^0} +\mu_{\Xi^-} + \mu_n = 2 \mu_\Lambda - \mu_p   \,\, , \,\,
\mu_\Lambda - \sqrt{3} \mu_{\Lambda\Sigma^0} = \mu_{\Xi^0} + \mu_n  \,\, . 
\eeq 
These are, in fact, in good agreement with the data. The other three
relations at this order are predictions for the deviations from the
Coleman--Glashow relations, the explicit forms can e.g. be found in 
refs.\cite{CP}\cite{jlms}. 
As shown below, these are the ones which pose problems.
 Before one can draw any conclusion about the
convergence or failure of chiral perturbation theory for the magnetic
moments, it is mandatory to get an estimate about the terms 
beyond ${\cal O}(q^3)$. 

\subsection{One--loop contribution at order $q^4$}

At order $q^4$ one has to include one loop graphs with insertions from 
${\cal L}_{MB}^{(2)}$ as well as additional tree graphs from
${\cal L}_{MB}^{(4)}$. Such an analysis has recently been performed for
the baryon masses and $\sigma$--terms \cite{bora}. 
Note that most of the ${\cal O}(q^4)$ contributions were included in
\cite{jlms}, which is based on an extended counting scheme. To see
this, one has to expand the diagrams with intermediate decuplet states
of \cite{jlms} in powers of the decuplet--octet splitting and retain
only the leading terms.  
Here, we attempt such a complete calculation for the magnetic moments
based on the conventional chiral counting with no explicit resonance fields.
The fourth order contribution to $\mu_B$ is given as follows (compare fig.~1)
\beq \label{mub4form}
\mu_B^{(4)} = \mu_B^{(4,c)} + \mu_B^{(4,d+e+f)} + \mu_B^{(4,g)} 
+ \mu_B^{(4,h+i)} + \mu_B^{(4,j)}  \, \, \, .
\eeq
The terms contributing to $\mu_B^{(4,c)}$ collect the counter term
(tree) contributions with exactly one insertion from the dimension
four effective Lagrangian. Here,
we consider the ones related to the explicit breaking of SU(3) due to the large
strange quark mass. These have the form\cite{bos}
\begin{eqnarray}\label{LMB4}
\nonumber
{\cal L}^{(4)}_{MB}  = 
& - & \frac{i\alpha_1}{4m} 
\langle \bar{B}[S^\mu,S^\nu]\left[[F^+_{\mu\nu},B],\chi^+\right] \rangle
-\: \frac{i\alpha_2}{4m} 
\langle \bar{B}[S^\mu,S^\nu]\left\{[F^+_{\mu\nu},B],\chi^+\right\}
\rangle \\
\nonumber
& - & \frac{i\alpha_3}{4m} 
\langle \bar{B}[S^\mu,S^\nu]\left[\{F^+_{\mu\nu},B\},\chi^+\right] \rangle
-\: \frac{i\alpha_4}{4m} 
\langle \bar{B}[S^\mu,S^\nu]\left\{\{F^+_{\mu\nu},B\},\chi^+\right\} \rangle\\
& - &\frac{i\beta_1}{4m} 
\langle \bar{B}[S^\mu,S^\nu]B \rangle \langle \chi^+ F^+_{\mu\nu} \rangle\:.
\end{eqnarray}
Here, $\chi^+$ is the spurion,
$\chi^+ = \,$diag(0,0,1), i.e. a factor $m_s$ has been
pulled out and absorbed in the low--energy constants $\alpha_{1,2,3,4}$ 
and $\beta_1$. The terms given in eq.(\ref{LMB4}) are of chiral dimension
four since $m_s = {\cal O}(q^2)$ and $F_{\mu \nu}  = {\cal
  O}(q^2)$. The LECs $\alpha_i$ and $\beta_1$ will be fitted in what
follows since the estimation from resonance saturation for such terms
is difficult and not yet under sufficient control as discussed in
detail in ref.\cite{bora}. Of course, in general these five terms
should be written in terms of the full quark matrix, but since $m_s
\gg m_d, m_u$, it is legitimate to neglect at this order the pionic 
contribution.  There are two more terms which could
contribute. These have the form 
\begin{equation} \label{LMB4mq}
{\cal L}_{MB}^{(4')} = 
-\frac{i}{4m} \, \tilde{b}_6^F \, \langle \chi_+ \rangle \,
\langle \bar{B} [S^\mu,
S^\nu][ F_{\mu \nu}^+, B] \rangle   
-\frac{i}{4m} \,\tilde{b}_6^D \, \langle \chi_+ \rangle \,
\langle \bar{B} [S^\mu,
S^\nu]\{ F_{\mu \nu}^+, B\} \rangle \,\, ,  
\end{equation}
where $\chi_+ = u^\dagger \chi u^\dagger + u \chi^\dagger u$,
$\chi =2B_0 {\cal M}$, with ${\cal M}$ the diagonal quark mass
matrix and $B_0 = |\langle 0|\bar q q|0\rangle|/F_\pi^2$ measures the strength
of the spontaneous symmetry violation.  We assume here $B \gg F_\pi$.
We have already written these
two terms in a way which makes obvious that the corresponding LECs
amount to quark mass renormalizations of $b_6^{D,F}$, i.e.
\beq \label{qmr}
b_6^{D,F} \to  b_6^{D,F} + \langle \chi_+ \rangle
 \, \tilde{b}_6^{D,F} \,\,\, .
\eeq 
Their contribution can therefore be absorbed in the values of the 
corresponding dimension two LECs. We note that the seven terms given
in eqs.(\ref{LMB4},\ref{LMB4mq}) have already been enumerated in 
\cite{jlms} (in other linear combinations).

The next type of graphs are the one--loop diagrams with exactly one 
insertion from the dimension two Lagrangian eq.({\ref{LMB2}),
i.e. they are proportional to the LECs $b_6^{D,F}$. For details on the
calculation of the wave--function renormalization (cf. fig.~1f), 
we refer to ref.\cite{bora}. These terms lead to large non--analytic
corrections of the form $m_q \ln m_q$, see also ref.\cite{jlms}. 
The various contributions for the baryon states are again 
relegated to appendix~A.
There are, however, three more different types of terms contributing. At
this order, we also have to consider double--derivative operators at
the meson--baryon vertex with the photon hooking on to the meson loop,
see fig.~1g. The corresponding terms of the dimension two Lagrangian
can be taken from ref.\cite{guido} 
\begin{equation} \label{LMB2p}
{\cal L}_{MB}^{(2')} = [S^\mu, S^\nu] \,  \biggl\{ b_9 \, 
 \langle \bar B \, u_\mu \rangle \langle u_\nu \, B \rangle   
+b_{10,11} \, \langle \bar{B} ( [u_\mu , u_\nu], B )_\pm \rangle
 \biggr\} \,\,\, ,   
\end{equation}
where $(..)_\pm$ means the anti-commutator and commutator,
respectively, with the corresponding LECs $b_{10}$ and $b_{11}$, in order.
Notice that since $u_\mu \sim \partial_\mu \phi + {\cal O}(\phi^2)$,
these terms lead to the structure $\partial_\mu \phi \partial_\nu
\phi$ mentioned in the introduction. This
contribution to the magnetic moments at quadratic order in the quark
masses has only been considered partly before. When one expands the one
loop graphs with intermediate decuplet states, one is lead to
operators of such type (plus an infinite tower of higher dimension
operators). In that case, however, one assumes that the corresponding 
LECs are entirely saturated by these decuplet excitations. Our
procedure is more general, we will estimate the values of the LECs
$b_{9,10,11}$ from resonance exchange as detailed in
section~3. This includes intermediate baryon excitations as well as
vector meson exchanges in the t--channel.
Furthermore, there are graphs with insertions from the part of  ${\cal
  L}_{MB}^{(2)}$ which is proportional to $1/m$ with {\it fixed}
coefficients, as depicted in fig.~1h+1i. The corresponding terms of
the Lagrangian read 
\begin{eqnarray}
{\cal L}_{MB}^{(2'')} & = &
\frac{1}{2m}\langle\bar{B}\left[v\cdot D,\big[v\cdot D,B\big]\right]\rangle
-\frac{1}{2m}\langle\bar{B}\left[D^\mu,\left[D_\mu,B\right]\right]\rangle
\nonumber \\[.5em]
& &
-\frac{i\,D}{4m}
\left(
\langle\bar{B}S_\mu\left[D^\mu,\left\{v\cdot u,B\right\}\right]\rangle
+\langle\bar{B}S_\mu\left\{v\cdot u,\left[D^\mu,B\right]\right\}\rangle
\right)\nonumber \\[.5em]
& &
-\frac{i\,F}{4m}
\left(
\langle\bar{B}S_\mu\left[D^\mu,\left[v\cdot u,B\right]\right]\rangle
+\langle\bar{B}S_\mu\left[v\cdot u,\left[D^\mu,B\right]\right]\rangle
\right)  \,\,\, . 
\end{eqnarray}
These terms are most economically constructed by use of the path integral
formalism developed for SU(2) in \cite{bkkm} and extended to SU(3) 
in \cite{guido}. Of course, they can also be constructed with no
recourse to the relativistic theory by means of reparametrization invariance
\cite{bclls}. These terms have not been calculated before. The explicit
expressions of these latter contributions to $\mu_B^{(4)}$ are also given
in appendix~A. Finally, there are self--energy  graphs (cf. fig.~1h)
with exactly one insertion from the scalar symmetry--breaking terms,
\begin{equation} \label{LMB2sb}
{\cal L}_{MB}^{(2''')} = 
b_D \, 
 \langle \bar B \, \{ \chi_+ , B \} \, \rangle 
+b_F \, \langle \bar{B} \, [\chi_+ , B ] \, \rangle \,\,\, .   
\end{equation} 
We call the diagrams belonging to this category class (j) graphs.
In principle, there is also the term $\sim b_0 \langle \bar B B
\rangle \langle \chi_+ \rangle$. For our purpose, however, it is
sufficient to lump its contribution together with the octet mass in
the chiral limit. The $b_0$ term can only be disentangled if one uses 
additional information like from the pion--nucleon $\sigma$--term. The
determination of the LECs $b_D$ and $b_F$ is relegated to the end of
the next section. Again, the explicit contributions to the fourth
order terms in the chiral expansion of the magnetic moments,
called $\mu_B^{(4,j)}$, can be found in appendix~A. We are now in the 
position to analyse the chiral expansion of  the magnetic moments.

\section{Estimation of some low--energy constants}

Since we do not take into account explicit resonance degrees of
freedom in our calculation, these show up indirectly in the values of
certain low--energy constants. In particular, it was shown in
ref.\cite{bkml2} that in the two--flavor case the LECs of the
dimension two Lagrangian can be understood in terms
of resonance exchange, largely due to the $\Delta$ and t--channel
vector meson excitations. In the three flavor case, the situation is
more difficult since no systematic investigation of low--energy
constants exists so far. In particular, it is difficult to get a 
handle on the symmetry breaking terms \cite{bora} like in eq.(\ref{LMB4}). 
We will therefore not estimate these from resonance exchange but
rather leave them as free (fit) parameters. 
On the other hand, the LECs
$b_{9,10,11}$ are related to Goldstone boson scattering off
baryons. In such a situation, resonance exchange can be used
easily to estimate the corresponding LECs.
We follow ref.\cite{bora} and use contributions from the
spin--3/2 decuplet, the isovector vector mesons and the spin-1/2 Roper
octet to fix these. The situation for the LECs  $b_6^{D,F}$ is
different. As shown in eq.(\ref{qmr}), these obtain a piece
proportional to the quark masses. Only the quark mass independent piece
could  be gotten from vector meson exchange and one would be left with the
fourth order pieces as fit parameters. We prefer not to perform this separation
and work with the total $b_6^{D,F}$ as fit parameters. Alternatively, one could
constrain the dimension two parts from resonance exchange and fit the
$\tilde{b}_6^{D,F}$. This is a general phenomenon which first appears at
order $q^4$ in the  baryon and at order $q^6$ in the meson sector, 
respectively.
Only if one is interested in the behaviour of certain
LECs as a function of the quark masses, one has to separate the pieces
from the various orders. We are not pursuing this issue here.

Consider now the LECs $b_{9,10,11}$. We treat the decuplet fields 
(collectively denoted by $\Delta$) 
relativistically and only at the last stage let the mass become very large.
The pertinent interaction Lagrangian reads
\beq
{\cal L}_{\Delta B \phi} = \frac{{\cal C}}{2} \biggl\{ \bar{\Delta}^{\mu,abc}
\Theta_{\mu\nu}(Z)(u^{\nu})_a^iB_b^j \epsilon_{cij} -  \bar{B}^b_i
(u^{^\mu})_j^a \Theta_{\mu\nu}(Z) {\Delta}^\nu_{abc} \epsilon^{cij} \biggr\}
\,\,,
\eeq
with $'a,b,c,i,j'$ SU(3) indices and the coupling constant ${\cal C}$
can be determined from the decays $\Delta \to B \pi$. $\Theta_{\mu\nu}(Z)$
is a Dirac operator,
\beq
\Theta_{\mu\nu}(Z) = g_{\mu\nu} - \biggl( Z + \frac{1}{2} \biggr) \gamma_\mu 
\gamma_\nu \,\,,
\eeq
and $Z$ is an off--shell parameter. With that, the meson--baryon scattering
tree graphs with an intermediate spin--3/2 decuplet state lead to 
\beq
b_9^\Delta = \frac{1}{2} \, I_\Delta \,\, , \, \, 
b_{10}^\Delta = \frac{1}{4} \, I_\Delta \,\, , \, \, 
b_{11}^\Delta = \frac{1}{12} \, I_\Delta \,\, , \, \,
\eeq
with 
\beq
I_\Delta = \frac{{\cal C}^2}{4} \biggl[ \frac{2}{3(m_\Delta -m)} +
\frac{m}{3m^2_\Delta} \bigl(1 + 4Z^2 + 4Z\bigl) + \frac{4Z}{m_\Delta} \biggr]
 \,\, ,
\eeq
and $m_\Delta = 1.38\,$GeV the average decuplet mass. In what
follows, we use  for the coupling
${\cal C} = 1.2 \ldots 1.5$ \cite{luwi} and for the off--shell parameter 
$Z=-0.3 \ldots 0$. The larger value corresponds to 
the static $\Delta$--isobar model 
and is thus in spirit closest to the study of ref.\cite{jlms} where a very
heavy explicit decuplet contribution was considered. The smaller value
stems from the study of the P-wave scattering volume $a_{33}$ \cite{armin}
treating the $\Delta$ as a relativistic field
(and also from pion photoproduction and nucleon polarizabilities).
We remind the reader that the pole piece due to the decuplet contribution
is unambiguous whereas the polynomial pieces depend on the off--shell
parameters. The latter dependence can always be absorbed in  local
contact terms as it is done here.
Similarly, the Roper octet (denoted $R$) contribution is derived from the
interaction Lagrangian
\beq
{\cal L}_{R B \phi} = \frac{D_R}{4} \langle \bar{R} \gamma^{\mu}\gamma_5
\{u_\mu, B\} \rangle +\frac{F_R}{4} \langle \bar{R} \gamma^{\mu}\gamma_5 
[u_\mu, B]\rangle + {\rm h.c.}
\,\,,
\eeq
with $D_R$ and $F_R$ axial--vector couplings that reduce in the two--flavor
case to $D_R + F_R = g_A \sqrt{\tilde R}$ (with $\sqrt{\tilde R}=0.53 \pm
0.04$ from the total width of the $N^* (1440)$ and $g_A = 1.33$ from the
Goldberger-Treiman relation). 
{}From the scattering graphs one finds (for more details, see \cite{bora})
\beq
b_9^R = -\frac{4}{3} \, D_R^2 \, I_R \,\, , \, \, 
b_{10}^R = 2 \,D_R \, F_R \, I_R \,\, , \, \, 
b_{11}^R = (D_R^2+F_R^2) \, I_R \,\, , 
\eeq
with
\beq
I_R = \frac{1}{16(m_R-m)} \quad .
\eeq
Here,  the average Roper octet mass is $m_R = 1.63\,$GeV and $F_R =
0.11$, $D_R = 0.60$ as determined in ref.\cite{bora}. Finally, there
is vector meson exchange. We use the vector-meson--pion
coupling ($V\phi \phi$) based on the antisymmetric tensor field 
notation of ref.\cite{reso} with $|G_V| = 69\,$MeV from 
$\Gamma (\rho \to \pi \pi )$. The universal vector meson
coupling $g_V$ is given by $g_V = g_{\rho \pi\pi} \, F_\pi^2 /
M_V^2$ and will be used in what follows. For the couplings of the
spin--1 fields to the baryons we follow \cite{bora2},
\beq
{\cal L}_{BBV} = K_{D/F} \, \langle \bar B \sigma^{\mu\nu}\,
(V_{\mu\nu} , B)_\pm \rangle \,\,\, ,
\eeq
where the two coupling constants $K_{D/F} $ are constrained by
\beq
K_D + K_F = -\frac{ g_{\rho NN} \, \kappa_V}{4 \, \sqrt{2} \, m}
\,\,\, , \eeq
by reduction to the two flavor case. Here, $\kappa_V = 6.1$ is the
tensor--to--vector coupling ratio of the $\rho$ and $V_{\mu\nu} =
\partial_\mu V_\nu - \partial_\nu V_\mu$ (making use of the equivalence
between the vector and tensor field formulations as proven in \cite{bora2}). 
To fix the relative strength of the  two tensor type couplings, 
we follow ref.\cite{thom}, $K_D/K_F = \alpha_M /(1 -\alpha_M)$ 
with $\alpha_M = 3/4$. Putting pieces together, we arrive at the contribution
from the t-channel isovector exchange,
\beq
b_9^V = 0 \,\, , \, \, 
b_{10}^V = \frac{\sqrt{2} \, G_V \, K_D}{ M_V}
= \frac{\kappa_V}{8m} \, {\alpha_M} \,\, , \, \, 
b_{11}^V =  \frac{\sqrt{2} \, G_V \, K_F}{M_V} 
= \frac{\kappa_V}{8m} \, {(1 - \alpha_M)} \,\, . 
\eeq

We now collect the various contributions and get  the following 
numerical values for these LECs for $Z = 0$ and ${\cal C} = 1.5$ 
(all numbers are in GeV$^{-1}$),
\beqa \label{b91011}
&& b_9 = b_9^\Delta + b_9^R + b_9^V = 0.87 - 0.06 + 0.00 = 0.81 \,\, , 
\nonumber \\
&& b_{10} = b_{10}^\Delta + b_{10}^R + b_{10}^V 
= 0.44 + 0.02 + 0.50 = 0.95 \,\, ,   \nonumber \\
&& b_{11} = b_{11}^\Delta + b_{11}^R + b_{11}^V 
= 0.15 + 0.05 + 0.17 = 0.36 \,\, ,
\eeqa
which will be used in the next section. We remark that the dimensionless
couplings $b_i' = 2m b_i$ are all of order one, i.e. they have "natural"
size. As expected from the study in \cite{jlms}, the decuplet contributes
much stronger than the Roper octet. However, whenever nonvanishing, the
vector meson exchange is also playing an important role. It is
important to stress that such type of operators were implicitely
included in the analysis of \cite{jlms}. This becomes apparent if one
expands the pertinent Feymnan diagrams in powers of $\Delta/ M_\pi$,
with $\Delta$ the average decuplet--octet mass splitting. This
generates (besides others) the contributions
$b_{9,10,11}^\Delta$. What is missing in \cite{jlms} are exactly the
vector meson terms shown to be of relevance in Eq.(\ref{b91011}). 
We remark that this way of treating the LECs induces a spurious
dependence on the scale of dimensional regularization $\lambda$. If
the LECs were to be fitted from data or resonance saturation would be
a perfect approximation, all observables would naturally be
independent of $\lambda$. To assess this sensitivity, we will later
vary $\lambda$ between $0.8$ to $1.2\,$GeV.

Finally, we have to determine the LECs $b_D$ and $b_F$,
cf. Eq.(\ref{LMB2sb}). Since these appear only in the one loop
graphs, it is sufficient to perform a fit to the baryon octet masses
at order $q^2$. To be precise, we identify the octet chiral limit mass, 
including the contribution from the term proportional to the  
LEC $b_0$, with the nucleon mass
and fit the two LECs from the splittings in the octet. This leads to
\beq
b_D = -0.192 \, {\rm GeV}^{-1} \,\,\, , \quad
b_F = -0.210  \, {\rm GeV}^{-1} \,\,\, .
\eeq
Two remarks are in order. First, we note that if one goes to higher
orders, the values for $b_{D,F}$ are usually different \cite{bkmz} \cite{bora}.
The procedure employed here is, however, consistent to 
the order we are working.
Second, the identification of the chiral limit mass 
supplemented by the $b_0$ contribution with the mass of
the nucleon is natural since all magnetic moments are measured in
units of nuclear magnetons, i.e. they refer to $1/m_N$.
We have also explored other options to treat these
parameters (the chiral limit mass, $b_0$, $b_D$, $b_F$)
but found the one described here most satisfactory.

\section{Results}

Consider now the chiral expansion at various orders.
A least square tree level fit based on the Lagrangian eq.(\ref{LMB2})
leads to $b_6^D = 2.39$ and $b_6^F =0.77$. The corresponding 
magnetic moments are given in table~1 in the column labelled ${\cal O}(q^2)$.
The empirical numbers (rounded to two digits) are reproduced to less than 20\%
with the exception of the $\Lambda$ and the $\Xi$'s. This is the typical
accuracy one expects from such a tree level SU(3) fit. Note, however,
that some of the Coleman--Glashow relations work much better, for example
$(\mu_n + \mu_p + \mu_{\Sigma^-}) / \mu_p  = 0$ \cite{CG} compared to the
empirical value of $-0.1$.

At the next order,
${\cal O}(q^3)$, one has large loop corrections (as already mentioned)
and thus needs to refit the two parameters $b_6^{D,F}$. With 
\begin{equation} \label{b6DF}
b_6^D = 5.29 \,\, , \quad b_6^F = 2.87 \,\, ,
\end{equation}
one obtains the values given in the third colum in table~1.
Clearly, the description of the magnetic moments did not improve, the average
deviation from the empirical values is still 25\%. The real problem, however,
is that the deviations from the Coleman--Glashow relations in most cases go
in the wrong direction, i.e. the gap to the experimental values of
 these ratios widens. To be precise, at order $q^2$ one has the three
relations $\mu_p - \mu_{\Sigma^+}=0$, $\mu_{\Xi^-} -\mu_{\Sigma^-}=0$
and $\mu_{\Xi^0} - \mu_n = 0$. At next order, these change to
$0.70$, $0.83$ and $1.67$ compared to the empirical (mean) values of
$0.37$, $0.51$ and $0.66$, respectively. Also, if one calculates the
mean deviation of the corrected Coleman--Glashow relations to the
empirical values, it increases to 44\% at ${\cal O}(q^3)$
from 22\% at leading order $q^2$.
This is essentially the source of the statement that the one--loop
chiral perturbation theory calculation fails to describe the baryon magnetic
moments or that one should decrease the rather large kaon loop contributions
as described above. 

{}From the previous considerations it is obvious that one has to go to
order $q^4$ to make a definite statement about the convergence and reliability
of the chiral expansion of the magnetic moments. Some of the non--analytic
terms at this order stemming from the insertion of the terms in eq.(\ref{LMB2})
in one loop graphs were considerd in \cite{jlms}. These lead to terms of the
type $m_q \ln m_q$, i.e. subleading non--analytic corrections. However, other
terms from the dimension two Lagrangian can contribute at ${\cal O}(q^4)$
as it is the case for the baryon masses as detailed in \cite{bora}. 
These together with the dimension four counter term contributions have
been calculated here. With seven free parameters, we can of course
fit  the measured magnetic moments (compare the fourth column in table~1).
The LECs take the values
\beqa \label{LECs}
&&b_6^D = 3.93 \, \, , \, \,  b_6^F = 3.01 \,\,\, , \nonumber \\
&&\alpha_1 = -0.45 \, , \, \alpha_2 = -0.06 \, , \, 
\alpha_3 = -0.47\, , \, \alpha_4 = 0.45 \, , \, \beta_1 = -0.74  \,\, .
\eeqa
The values of $b_6^{D,F}$ are changed compared to the
${\cal O}(q^3)$ fit which indicates that  the quark mass
dependent corrections of order $q^4$ are not small.
 The numerical values of the dimension four LECs  
are of the size expected from naive dimensional analysis since they should be
suppressed  by a factor $m_s /
\Lambda_\chi \sim 1/6$ compared to the dimension two couplings
(with $\Lambda_\chi \simeq 1\,$GeV the scale of
chiral symmetry breaking and $m_s (\Lambda_\chi ) = 175\,$MeV). 
Clearly, the
predictive power of our approach is not very large, only the $\Lambda \Sigma^0$
transition moment is predicted. Our intention is, however, more modest. We want
to study the convergence of the chiral expansion of the magnetic moments. It
appears to be under control (we show the results for $Z=0$ and ${\cal C}=1.5$), 
\begin{eqnarray} \label{conv}
\begin{array}{llrllr}
\mu_p               & = & 4.31 & (1 - 0.51 + 0.16) & = & 2.79 \, , \\
\mu_n               & = & -2.61 & (1 - 0.32 + 0.05) & = & -1.91 \, , \\
\mu_{\Sigma^+}      & = & 4.31 & (1 - 0.64 + 0.21) & = & 2.46 \, , \\
\mu_{\Sigma^-}      & = & -1.70 & (1 - 0.36 + 0.05) & = & -1.16 \, , \\
\mu_{\Sigma^0}      & = & 1.31 & (1 - 0.82 + 0.32) & = & 0.65 \, , \\
\mu_\Lambda         & = & -1.31 & (1 - 0.82 + 0.29) & = & -0.61 \, , \\
\mu_{\Xi^0}         & = & -2.62 & (1 - 0.84 + 0.32) & = & -1.25 \, , \\
\mu_{\Xi^-}         & = & -1.70 & (1 - 0.76 + 0.15) & = & -0.65 \, , \\
\mu_{\Lambda\Sigma} & = & 2.28 & (1 - 0.50 + 0.13) & = & 1.42 \, . \\
\end{array} 
\end{eqnarray}
We have set here the scale of dimensional regularization $\lambda =
800\,$ MeV. In all cases the ${\cal O}(q^4)$ contribution is smaller than the one 
from ${\cal O}(q^3)$ by at least a factor of two, in most cases even by 
a factor of three. 
Like in the case of the baryon masses 
\cite{liz} \cite{bora}, we find sizeable
cancellations between the leading and next--to--leading order terms
making a {\it precise} calculation of the ${\cal O}(q^4)$ terms absolutely
necessary. For the $\Sigma^0$, the $\Lambda$ and the $\Xi^0$ these
cancellations are so effective that the fourth order contribution appears
at first sight  fairly large.     The result for
the $\Sigma^0$ is in agreement with isospin symmetry (as it should) and the
prediction for the
transition moment $\mu_{\Lambda \Sigma^0}= 1.42$ is within 1.5
standard deviations of the recent
lattice calculation of \cite{lwd}, $\mu_{\Lambda \Sigma^0}= 1.54 \pm 0.09$.
It is somewhat smaller in magnitude than the PDG value, 
$\mu_{\Lambda \Sigma^0}= \pm (1.61 \pm 0.08)$ but within 2.5 
standard deviations. To make a more realistic prediction for 
$\mu_{\Sigma^0}$, one
would have to account for effects of the order $m_u - m_d$ and include
virtual photons. This goes beyond the scope of this work.

The various contribution to the fourth order term $\mu_B^{(4)}$ are
separately given in table~2  (for $Z=0$ and ${\cal C}=1.5$). 
Individually, some of these contributions
are rather large but in all cases sizeable cancellations appear that
make the sum relatively small. In particular, we stress that the
class~g diagrams, which were only considered to some extent before,
 give in some cases even the
{\it largest} contribution. Furthermore, the corrections due to $1/m$ 
insertions with fixed coefficients are quite sizeable
in certain cases like the $\Sigma^+$. The terms due
to scalar symmetry breaking insertions $\sim b_{D,F}$ are in
general small, with the exception of the $\Sigma^+$ and the
$\Sigma^-$.  This underlines the statement
that it is mandatory to take {\it all} terms at a given order in the 
chiral expansion. 

It is also instructive to disentangle in the fourth order pieces
the various contributions from the decuplet, the Roper octet and the
vector meson contribution. In all but one cases, the decuplet
and vector meson contributions are of comparable size and much bigger
than the one from the Roper octet. Two typical cases are the proton
and the neutron (for $Z = 0$ and ${\cal C} = 1.5$),
\beqa
\begin{array}{lllrlrlr}
p &:& \quad \mu^{(4,\Delta)}_p = & -0.42 \,\, , & 
\mu^{(4,V)}_p = & -0.27 \,\, , & \mu^{(4,R)}_p = & -0.04 \, \, \, ,  \\
& & & & \\
n &:& \quad \mu^{(4,\Delta)}_n = &  0.23 \,\, , & 
\mu^{(4,V)}_n = &  0.27 \,\, , & \mu^{(4,R)}_n = & 0.0 \, \, \, . \\
\end{array}
\eeqa
This underlines the previously made statement that the t--channel
vector meson excitations play an important role in determimning the
strength of the insertions $\sim b_{9,10,11}$ and that it is not
sufficient to only take the decuplet to saturate these LECs.

If one varies the parameters $Z$ and ${\cal C}$ within the bounds
discussed in section 3, the trends discussed above are the same.
There is some reshuffling between the various fourth order 
contributions, but the sum of them stays approximately the same.
Reducing the values for $Z$ and ${\cal C}$ tends to increase the
fourth order contribution by some percent. For the proton, the
relative fourth order contribution varies between $0.16$ and $0.20$
for the ranges of $Z$ and ${\cal C}$ given above. We note that within
these ranges,  the
prediction for the transition magnetic moment is stable
within $0.01\,\mu_N$. For orientation, we give in table~3
the values of the dimension two and four LECs for various
input values of $Z$ and ${\cal C}$. We remark that the relation
between certain magnetic moments derived in \cite{jlms} does not hold in the
complete $q^4$ calculation presented here. 

Let us now discuss the scale dependence of the results presented so
far. As already explained, our resonance saturation estimate leads to
some spurious dependence on $\lambda$. Typically, one would vary this
scale between the mass of the $\rho$ and  the mass of the $\Delta$.
In Fig.~2, we show the relative third and fourth order contributions
(the second order term is normalized to give one) as a function of
$\lambda$ for the proton (which is a representative state). As $\lambda$
increases, the fourth order terms become more and more
important. However, they stay well below the third order
contributions. This behaviour is expected since the contribution from
the chiral logarithms grows as  $\lambda$ increases. For example,
$\ln(M_K / 0.8) = -0.47$ where as $\ln(M_K / 1.2) = -0.88$. This increase
is only balanced in part by the readjustment of the LECs. In particular,
the LECs $b_6^D$ and $b_6^F$ become smaller and thus the leading second
order contribution decreases which in turn makes the relative third
and fourth order contributions grow.

Finally, we remark that there might be potentially large corrections at
higher orders. Not having performed a complete calculation at ${\cal
  O}(q^5)$, the following remarks should be taken with caution. We
simply note that using the scalar symmetry breaking LECs $b_D$ and
$b_F$ as determined from a $q^3$ \cite{bkmz} or $q^4$ \cite{bora} fit
leads to very large corrections for the proton from the class (j)
diagrams which in turn makes the fit values for the $\alpha_i$ and
$\beta_1$ unnaturally large. However, there are many other
contributions at that order and only a complete calculation beyond the
one presented here could unravel the true significance of these effects.

\section{Summary and conclusions}

In this paper, we have considered the chiral expansion of the
octet baryon magnetic moments at next--to--next--to--leading
order ${\cal O}(q^4)$. Such a study is mandated by the fact
that while the leading order $(q^2)$ relations originally derived
by Coleman and Glashow work quite well, considering the leading
and next--to--leading non--analytic corrections of the type
$m_q^{1/2}$ and $m_q \ln m_q $ does not improve the description
of the magnetic moments and has even some times been advocated as
a breakdown of baryon chiral perturbation theory for three flavors.
Our study extends the one of ref.\cite{jlms} in that we have included
{\it all} terms at this order.\footnote{We stress again that we use
  the conventional power counting and do not include the decuplet in
  the effective field theory. Our results  can be matched onto the ones 
  of \cite{jlms} if one expands in powers of $(m_\Delta -m_N) /M_\pi$. 
  How this procedure works in case of the baryon masses is shown in
  ref.\cite{bkmz}. We furthermore remark that some
  $1/m$ corrections have not been included in \cite{jlms}.} 
In particular, we have shown that
the double--derivative meson--baryon couplings generated in \cite{jlms} by
expanding loop graphs with intermediate decuplet states receive
important corrections from vector meson exchange. In total, we have
to deal with seven low--energy constants related to symmetry breaking
in the presence of the photon field, two LECs related to scalar
symmetry breaking
and three which appear in meson--baryon scattering. The latter
can be estimated with some confidence from resonance exchange, in
our case from decuplet and vector meson excitation. This means in
particular that we do not 
include the spin--3/2 fields as dynamical degrees of freedom in
the effective theory. As stressed in ref.\cite{bora}, a reliable
estimation of the LECs related to symmetry breaking is very difficult.
We have therefore kept the ones related to symmetry breaking
in the presence of electromagnetic fields as free parameters and 
are thus able to
fit the observed magnetic moments. The two LECs related to the scalar
sector could be determined from a best fit to the octet mass
splittings at order $q^2$. Our main purpose was to study the
convergence of the chiral expansion. We find that with the exception
of the $\Lambda$, $\Sigma^0$ and $\Xi^0$, there appears to be
good convergence. In the three exceptional cases, the leading and
next--to--leading order terms cancel so that the ${\cal O}(q^4)$
contribution is somewhat enlarged. As discussed in detail, there appear
large cancellations between the various classes of diagrams at
order $q^4$. In particular, in most cases the  contribution
from the one loop graphs with a double derivative dimension two
vertex is substantial and it is thus mandatory to include
these terms as precisely as possible. The novel ingredient in our
calculation is the substantial contribution of t--channel vector meson
exchange to the LECs pertinent to these operators.
Only the complete order $q^4$ calculation performed
here can be used to explore the convergence of the chiral expansion.
The results  obtained are stable against variations
in the input parameters, in particular to the ones related to the
decuplet sector. We can also predict the $\Lambda \Sigma^0$ transition
moment, $\mu_{\Lambda \Sigma^0} = (1.42 \pm 0.01) \mu_N$ in fair 
agreement with recent lattice gauge theory results. Notice that the
theoretical error given for $\mu_{\Lambda \Sigma^0}$ refers to the
order we have been working and does not account for the uncertainty
related to higher orders not yet calculated.  What still has to
be gained is a deeper theoretical understanding of the numerical
values of the LECs related to symmetry breaking. 
Work along these lines is under way.

\vfill

\section*{Acknowledgements}

We thank V. Bernard, B. Borasoy, N. Kaiser and G. M\"uller for useful comments.
One of us (UGM) thanks Prof. H. Toki and the Nuclear Theory group
at RCNP Osaka for hospitality during a stay when part of this work
was done. We are grateful to Phuoc Ha for pointing out an inconsistency in the
first version of the manuscript.

\vspace{1.5cm}

\appendix
\section{Formulae for the magnetic moments}
\def\theequation{\Alph{section}.\arabic{equation}}
\setcounter{equation}{0}

In this appendix, we collect the explicit formulae for the
chiral expansion of the baryon magnetic moments to order
$q^2$, $q^3$ and $q^4$, respectively. The latter are grouped according
to the various contributions discussed in the main text. 

\medskip

\noindent  \underline{Order $q^2$ (compare fig.~1a)}

\begin{equation}
\mu_B^{(2)} = \alpha_B^D \, b_6^D + \alpha_B^F \, b_6^F \,\,\, ,
\end{equation}
with 
\begin{eqnarray}
&& \alpha_p^D = 1/3 \,\, , \,\, \alpha_p^F = 1 \,\, ,  \,\,
\alpha_n^D = -2/3 \,\, , \,\, \alpha_n^F = 0 \,\, , \, \,
\alpha_{\Lambda}^D = -1/3 \,\, , \,\, \alpha_{\Lambda}^F = 0 \,\, ,
\nonumber \\
&& \alpha_{\Sigma^+}^D = 1/3 \,\, , \,\, \alpha_{\Sigma^+}^F =1 \,\, ,  \,\,
\alpha_{\Sigma^-}^D = 1/3 \,\, , \,\, \alpha_{\Sigma^-}^F = -1 \,\, ,
\alpha_{\Sigma^0}^D = 1/3 \,\, , \,\, \alpha_{\Sigma^0}^F = 0 \,\, , \\
&&\alpha_{\Xi^-}^D = 1/3 \,\, , \,\, \alpha_{\Sigma^-}^F = -1 \,\, , \,\,
\alpha_{\Xi^0}^D = -2/3 \,\, , \,\, \alpha_{\Xi^0}^F = 0 \,\, , \,\,
\alpha_{\Lambda\Sigma^0}^D = 1/\sqrt{3} \,\, , \,\, 
\alpha_{\Lambda\Sigma^0}^F = 0 \,\, . \nonumber 
\end{eqnarray}
  
\bigskip

\noindent \underline{Order $q^3$ (compare fig.~1b)}

\begin{equation}
\mu_B^{(3)} = \beta^\pi \, \frac{M_\pi\,m}{8\pi F_\pi^2} + 
\beta^K \,\frac{M_K\,m}{8\pi F_K^2} \,  \,\,\, ,
\end{equation}
with 
\begin{eqnarray}
&& \beta_p^\pi = -(D+F)^2 \,\, , \,\, \beta_p^K = - 2/3 \, (D^2+3F^2) \,\, ,  \,\,
\beta_n^\pi = (D+F)^2 \,\, , \,\, \beta_n^K = - (D-F)^2 \,\, , \, \,
\nonumber \\
&&
\beta_{\Lambda}^\pi = 0 \,\, , \,\, \beta_{\Lambda}^K = 2DF \,\, ,
\beta_{\Sigma^+}^\pi = - 2/3 \, (D^2+3F^2) \,\, , \,\,
\beta_{\Sigma^+}^K = -(D+F)^2 \,\, ,  \,\,
\nonumber \\
&&
\beta_{\Sigma^-}^\pi = 2/3 \, (D^2+3F^2) \,\, , \,\, \beta_{\Sigma^-}^K = (D-F)^2\,\, ,
\beta_{\Sigma^0}^\pi = 0 \,\, , \,\, \beta_{\Sigma^0}^K = -2DF \,\, , \\
&&\beta_{\Xi^-}^\pi = (D+F)^2 \,\, , \,\, \beta_{\Sigma^-}^K = 2/3 \, (D^2+3F^2) \,\, , \,\,
\beta_{\Xi^0}^\pi = - (D-F)^2 \,\, , \,\, \beta_{\Xi^0}^K = (D+F)^2 \,\, , \,\,
\nonumber \\
&&
\beta_{\Lambda\Sigma^0}^\pi = -4/\sqrt{3} \, DF\,\, , \,\,
\beta_{\Lambda\Sigma^0}^K =  -2/\sqrt{3} \, DF\,\, . \nonumber 
\end{eqnarray}

\bigskip

\noindent \underline{Order $q^4$ (compare fig.~1c - 1i)}

\begin{equation}
\mu_B^{(4)} =  \mu_B^{(4,c)} + \mu_B^{(4,d+e+f)} + \mu_B^{(4,g)} 
+  \mu_B^{(4,h+i)}  
\end{equation}
with
\begin{eqnarray}
&&
\mu^{(4,c)}_p = \alpha_1 + \alpha_2 + 1/3\,\alpha_3 + 1/3\,\alpha_4 - 1/3\,\beta_1 \,\, , \,\, 
\mu^{(4,c)}_n = - 2/3\,\alpha_3 - 2/3\,\alpha_4 - 1/3\,\beta_1 \,\, , \nonumber\\
&&
\mu^{(4,c)}_\Lambda = -8/9 \, \alpha_4 - 1/3 \, \beta_1 \,\, , \,\,
\mu^{(4,c)}_{\Sigma^+} =
\mu^{(4,c)}_{\Sigma^-} = 
\mu^{(4,c)}_{\Sigma^0} = - 1/3 \, \beta_1 \,\, , \,\,
\mu^{(4,c)}_{\Lambda\Sigma^0} = 0 \,\, , \\
&&
\mu^{(4,c)}_{\Xi^-} = \alpha_1 - \alpha_2 - 1/3\,\alpha_3 + 1/3\,\alpha_4 - 1/3\,\beta_1 \,\, , \,\, 
\mu^{(4,c)}_{\Xi^0} = 2/3\,\alpha_3 - 2/3\,
\alpha_4 - 1/3\,\beta_1 \,\, . \nonumber
\end{eqnarray}
and
\begin{eqnarray}
\mu_B^{(4,d+e+f)} & = & 
\delta^\pi_B \, \frac{M_\pi^2}{16\pi^2 F_\pi^2} \, \ln\frac{M_\pi}{\lambda}
+\delta^K_B \, \frac{M_K^2}{16\pi^2 F_K^2} \, \ln\frac{M_K}{\lambda} 
\nonumber
+\sum_{X=\pi,\,K,\,\eta}\eta^X_B\,\frac{M_X^2}{16\pi^2
  F_X^2}\,\ln\frac{M_X}{\lambda}
\\
& &
+\left(
\sum_{X=\pi,\,K,\,\eta}\phi^X_B \, \frac{M_X^2}{8\pi^2 F_X^2}\,\left(3\ln\frac{M_X}{\lambda}+1\right)
\right)
\mu^{(2)}_B
\end{eqnarray}
with
\begin{eqnarray}
&&
\delta^\pi_p = - b_6^F - b_6^D \,\, , \,\,
\delta^K_p = - 2 \, b_6^F \,\, , \,\ 
\delta^\pi_n =  b_6^D + b_6^F \,\, , \,\
\delta^K_n = b_6^D - b_6^F \,\, , \,\
\delta^\pi_\lambda = 0 \,\, , \,\
\delta^K_\lambda = b_6^D \,\, , \nonumber \\
&&
\delta^\pi_{\Sigma^+} = - 2 \, b_6^F \,\, , \,\
\delta^K_{\Sigma^+} = - b_6^D - b_6^F \,\, , \,\
\delta^\pi_{\Sigma^-} = 2 \, b_6^F \,\, , \,\
\delta^K_ {\Sigma^-} = b_6^F - b_6^D \,\, , \,\
\delta^\pi_{\Sigma^0} = 0 \,\, , \,\
\delta^K_{\Sigma^0} = b_6^D \,\, , \nonumber \\
&&
\delta^\pi_{\Xi^-} = b_6^F - b_6^D  \,\, , \,\
\delta^K_{\Xi^-} = 2 \, b_6^F \,\, , \,\
\delta^\pi_{\Xi^0} =  b_6^D - b_6^F \,\, , \,\
\delta^K_{\Xi^0} = b_6^D + b_6^F \,\, , \\
&&
\delta^\pi_{\Lambda\Sigma^0} = -2/\sqrt{3} \, b_6^D \,\, , \,\,
\delta^K_{\Lambda\Sigma^0} = -1/\sqrt{3} \, b_6^D \,\, . \nonumber
\end{eqnarray}
and
\begin{eqnarray}
&& \nonumber
\eta^\pi_p = (D+F)^2/2 (b_6^D - b_6^F) \,\, ,\,\,
\eta^K_p = -(D^2/9-2DF+F^2) b_6^D - (D-F)^2 b_6^F \,\, ,\\ 
&& \nonumber
\eta^\eta_p = - (D-3F)^2/18 (b_6^D + 3b_6^F) \,\, ,\,\,
\eta^\pi_n = - (D+F)^2 b_6^F \,\, ,\\
&& \nonumber
\eta^K_n = (-7/9D^2+2/3DF+F^2) b_6^D + (D-F)^2  b_6^F \,\, ,\,\,
\eta^\eta_n = 1/9\,(D-3F)^2\, b_6^D \,\, ,\\
&& \nonumber
\eta^\pi_\Lambda = -2/3\,D^2 b_6^D \,\, ,\,\,
\eta^K_\Lambda = 1/9\,(D^2+9F^2)\, b_6^D - 2DF  b_6^F \,\, ,\,\,
\eta^\eta_\Lambda = 2/9\,D^2\, b_6^D \,\, ,\\
&& \nonumber
\eta^\pi_{\Sigma^+} = 2/9\,(D^2+6DF-6F^2)\,b_6^D - 2F^2 b_6^F \,\, ,\\
&& \nonumber
\eta^K_{\Sigma^+} = 1/3\,(D^2+6DF+F^2) b_6^D - (D-F)^2  b_6^F \,\, ,\,\,
\eta^\eta_{\Sigma^+} = -2/9\,D^2\,(b_6^D + 3b_6^F) \,\, ,\\
&& \nonumber
\eta^\pi_{\Sigma^-} = 2/9\,(D^2-6DF-6F^2)\,b_6^D + 2F^2 b_6^F \,\, ,\\
&& \nonumber
\eta^K_{\Sigma^-} = 1/3\,(D^2-6DF+F^2) b_6^D + (D+F)^2  b_6^F \,\, ,\,\,
\eta^\eta_{\Sigma^-} = -2/9\,D^2\,(b_6^D - 3b_6^F) \,\, ,\\
&& \nonumber
\eta^\pi_{\Sigma^0} =  2/9\,(D^2-6F^2)\,b_6^D  \,\, ,\,\, 
\eta^K_{\Sigma^0} = 1/3\,(D^2+F^2) b_6^D + 2DF b_6^F   \,\, ,\\
&& \nonumber
\eta^\eta_{\Sigma^0} = -2/9\,D^2\,b_6^D \,\, ,\,\,
\eta^\pi_{\Xi^-} = 1/2\,(D-F)^2\,(b_6^D + b_6^F) \,\, ,\\
&& \nonumber
\eta^K_{\Xi^-} = (-1/9D^2-2DF-F^2)\,b_6^D + (D+F)^2\,b_6^F \,\, ,\\
&& \nonumber
\eta^\eta_{\Xi^-} = -1/18\,(D+3F)^2(b_6^D - 3 b_6^F) \,\, ,\,\,
\eta^\pi_{\Xi^0} = (D-F)^2 b_6^F \,\, ,\\
&& \nonumber
\eta^K_{\Xi^0} = (-7/9D^2-2/3DF+F^2) b_6^D - (D+F)^2  b_6^F \,\, ,\,\,
\eta^\eta_{\Xi^0} = 1/9\,(D+3F)^2 b_6^D \,\, ,\\
&& \nonumber
\eta^\pi_{\Lambda\Sigma^0} = 2/3/\sqrt{3}\,(6DF b_6^F - D^2 b_6^D) \,\, ,\,\,
\eta^K_{\Lambda\Sigma^0} = 1/\sqrt{3}\,(D^2-3F^2) b_6^D + 2/\sqrt{3}\,DF  b_6^F
\,\, ,\\
&& 
\eta^\eta_{\Lambda\Sigma^0} = 2/(3\sqrt{3})\,D^2\,b_6^D \,\, .
\end{eqnarray}
The wavefunction renormalization coefficients are invariant under
SU(2) transformations, therefore we give only the values for the different
multiplets:
\begin{eqnarray}
&& \nonumber
\phi_N^\pi = 3/4\,(D+F)^2 \,\, ,\,\,
\phi_N^K = 5/6D^2-DF+3/2F^2 \,\, ,\,\,
\phi_N^\eta = 1/12(D-3F)^2 \,\, ,\\
&& \nonumber
\phi_\Sigma^\pi = 1/3D^2+2F^2 \,\, ,\,\,
\phi_\Sigma^K = D^2+F^2 \,\, ,\,\,
\phi_\Sigma^\eta = 1/3 D^2 \,\, ,\\
&&
\phi_\Lambda^\pi = D^2 \,\, ,\,\,
\phi_\Lambda^K = 1/3D^2+3F^2 \,\, ,\,\,
\phi_\Lambda^\eta = 1/3 D^2 \,\, ,\\
&& \nonumber
\phi_\Xi^\pi = 3/4(D-F)^2 \,\, ,\,\,
\phi_\Xi^K = 5/6D^2+DF+3/2F^2 \,\, ,\,\,
\phi_\Xi^\eta = 1/12(D+3F)^2 \,\, .
\end{eqnarray}
and
\begin{equation}
\mu_B^{(4,g)} = \gamma^\pi_B \, \frac{m\,M_\pi^2}{4\pi^2 F_\pi^2} \,
\ln\frac{M_\pi}{\lambda}
+ \gamma^K_B \, \frac{m\,M_K^2}{4\pi^2 F_K^2} \,
\ln\frac{M_K}{\lambda}
\end{equation}
with
\begin{eqnarray}
&&
\gamma^\pi_p = 2 \, b_{10} + 2 \, \tilde{b}_{11} \,\, , \,\,
\gamma^K_p = b_9 + 4 \,\tilde{b}_{11} \,\, , \,\,
\gamma^\pi_n = - 2 \, b_{10} - 2 \,\tilde{b}_{11} \,\, , \,\,
\gamma^K_n = - 2 \, b_{10} + 2 \,\tilde{b}_{11} \,\, , \nonumber\\
&&
\gamma^\pi_\Lambda = 0 \,\, , \,\, 
\gamma^K_\Lambda = - 2 \, b_{10} \,\, , \,\,
\gamma^\pi_{\Sigma^+} = b_9 + 4 \,\tilde{b}_{11} \,\, , \,\,
\gamma^K_{\Sigma^+} = 2 \, b_{10} + 2 \,\tilde{b}_{11} \,\, , \nonumber \\
&&
\gamma^\pi_{\Sigma^-} = - b_9 - 4 \,\tilde{b}_{11} \,\, , \,\,
\gamma^K_{\Sigma^-} = 2 \, b_{10} - 2 \,\tilde{b}_{11} \,\, , \,\,
\gamma^\pi_{\Sigma^0} = 0 \,\, , \,\,
\gamma^K_{\Sigma^0} = 2 \, b_{10} \,\, , \\
&&
\gamma^\pi_{\Xi^-} = 2 \, b_{10} - 2 \,\tilde{b}_{11} \,\, , \,\,
\gamma^K_{\Xi^-} = - b_9 - 4 \,\tilde{b}_{11} \,\, , \,\ 
\gamma^\pi_{\Xi^0} = - 2 \, b_{10} + 2 \,\tilde{b}_{11} \,\, , \nonumber \\
&&
\gamma^K_{\Xi^0} =  - 2 \, b_{10} - 2 \,\tilde{b}_{11} \,\, , \,\,
\gamma^\pi_{\Lambda\Sigma^0} = 4/\sqrt{3} \, b_{10} \,\, , \,\, 
\gamma^K_{\Lambda\Sigma^0} = 2/\sqrt{3} \, b_{10} \,\, , \nonumber
\end{eqnarray}
and
\beq
\tilde{b}_{11} = b_{11} + {1 \over 8m} \,\,\, .
\eeq
For the graphs with fixed coefficients $\sim 1/m$, one gets
\begin{equation}
\mu_B^{(4,h+i)} = \beta^\pi_B  \, \frac{3\,M_\pi^2}{8\pi^2
  F_\pi^2}\,\ln\frac{M_\pi}{\lambda}
+ \beta^K_B \, \frac{3\,M_K^2}{8\pi^2 F_K^2}\,\ln\frac{M_K}{\lambda} \, \,\, .
\end{equation}
Finally, the corrections due to the baryon mass differences read:
\begin{equation}
\mu_B^{(4,j)} = 
-\theta_B^\pi\, \frac{m}{2\pi^2 F_\pi^2} 
\left(1+\ln\frac{M_\pi}{\lambda}\right)
-
\theta_B^K\, \frac{m}{2\pi^2 F_K^2} 
\left(1+\ln\frac{M_K}{\lambda}\right)
\end{equation}
with
\begin{eqnarray}
&&
\theta_p^\pi =  (D+F)^2 (M_K^2\,b_D + (M_\pi^2-M_K^2)\,b_F)\,\,,\,\,
\nonumber
\\
&&
\theta_p^K = 1/6\,((3F+D)^2 M_\eta^2 + 3(D-F)^2 M_\pi^2)\,b_D\,\,,
\nonumber
\\
&&
\theta_n^\pi = -(D+F)^2 (M_K^2\,b_D + (M_\pi^2-M_K^2)\,b_F)\,\,,\,\,
\theta_n^K = (D-F)^2 M_\pi^2\,b_D\,\,,
\nonumber
\\
&&
\theta_\Lambda^\pi = 0\,\,,\,\,
\theta_\Lambda^K = -2 DF M_K^2\,b_D + 1/3\,(9F^2+D^2)(M_K^2-M_\pi^2)\,b_F\,\,,
\nonumber
\\
&&
\theta_{\Sigma^+}^\pi = 2/3 (3F^2 M_\pi^2+D^2 M_\eta^2)\,b_D \,\,,\,\,
\theta_{\Sigma^+}^K = (D+F)^2 (M_K^2\,b_D - (M_\pi^2-M_K^2)\,b_F)\,\,,
\nonumber
\\
&&
\theta_{\Sigma^-}^\pi = -2/3 (3F^2 M_\pi^2+D^2 M_\eta^2)\,b_D \,\,,\,\,
\theta_{\Sigma^-}^K = -(D-F)^2 (M_K^2\,b_D + (M_\pi^2-M_K^2)\,b_F)\,\,,
\nonumber
\\
&&
\theta_{\Sigma^0}^\pi = 0\,\,,\,\,
\theta_{\Sigma^0}^K = 2 DF M_K^2\,b_D + (F^2+D^2)(M_K^2-M_\pi^2)\,b_F\,\,,
\nonumber
\\
&&
\theta_{\Xi^-}^\pi = -(D-F)^2 (M_K^2\,b_D - (M_\pi^2-M_K^2)\,b_F)\,\,,\,\,
\nonumber
\\
&&
\theta_{\Xi^-}^K = -1/6\,((3F^2+D^2)M_\eta^2+3(D+F)^2 M_\pi^2)\,b_D \,\,,
\nonumber
\\
&&
\theta_{\Xi^0}^\pi = (D-F)^2 (M_K^2\,b_D + (M_K^2-M_\pi^2)\,b_F)\,\,,\,\,
\theta_{\Xi^0}^K = -(D-F)^2 M_\pi^2\,b_D\,\,,
\nonumber
\\
&&
\theta_{\Lambda\Sigma^0}^\pi = 4/\sqrt{3}\,DF M_\pi^2\,b_D\,\,,\,\,
\nonumber
\\
&&
\theta_{\Lambda\Sigma^0}^K = 2/\sqrt{3}\, DF M_K^2\,b_D + 1/\sqrt{3}\, (3F^2-D^2)(M_K^2-M_\pi^2)\,b_F\,\,.
\end{eqnarray}
\vfill\eject


\vfill\eject


\noindent{\Large {\bf Tables}}


\renewcommand{\arraystretch}{1.1}


\begin{center}

\begin{tabular}{|l|r|r|r|r|}
    \hline
    $\mu_B$ & ${\cal O}(q^2)$ &  ${\cal O}(q^3)$ &  ${\cal O}(q^4)$ & Exp. \\
    \hline
    p                 &  2.56   &  2.94   &  2.79   & $2.793 \pm 0.000$ \\    
    n                 & $-$1.60 & $-$2.49 & $-$1.91 & $-1.913 \pm 0.000$ \\
    $\Sigma^+$        &  2.56   &  2.24   &  2.46   & $2.458 \pm 0.010$ \\
    $\Sigma^-$        & $-$0.97 & $-$1.35 & $-$1.16 & $-1.160 \pm 0.025$ \\
    $\Sigma^0$        &  0.80   &  0.44   &  0.65   &  $-$   \\
    $\Lambda$         & $-$0.80 & $-$0.44 & $-$0.61 & $-0.613 \pm 0.004$ \\
    $\Xi^0$           & $-$1.60 & $-$0.82 & $-$1.25 & $-1.250 \pm 0.014$ \\
    $\Xi^-$           & $-$0.97 & $-$0.52 & $-$0.65 & $-0.651 \pm 0.003$ \\
    $\Lambda\Sigma^0$ &  1.38   &  1.65   &  1.42   & $\pm1.61 \pm 0.08$\\
    \hline
  \end{tabular}

\bigskip 

Table 1: Magnetic moments (in $\mu_N$) calculated to various orders.


\end{center}

\bigskip

\renewcommand{\arraystretch}{1.1}


\begin{center}

\begin{tabular}{|l|r|r|r|r|r|r|}
    \hline
    $\mu_B^{(4)}$ & c &  d+e+f &   g  & h+i & j & sum \\
    \hline
    p           &  $-$0.27  &  1.27  &  $-$0.88 & 0.54 & 0.01 & 0.67 \\    
    n           &  0.26 & $-$0.65 & 0.47 & $-$0.21 & $-$0.01 & $-$0.14 \\
    $\Sigma^+$  &  0.25  &  1.19   &  $-$0.89 & 0.67 & $-$0.30 & 0.91 \\
    $\Sigma^-$  &  0.25 & $-$0.56 & 0.06 & $-$0.15 & 0.32 & $-$0.08 \\
    $\Sigma^0$  &  0.25 & 0.31 &$-$0.42 & 0.26 & 0.01 & 0.42   \\
    $\Lambda$   & $-$0.15 & $-$0.39 & 0.42 & $-$0.26 & 0.0 & $-$0.38 \\
    $\Xi^0$     & $-$0.36 & $-$0.57 & 0.72 & $-$0.53 & $-$0.08 & $-$0.84 \\
    $\Xi^-$     & 0.16 & $-$0.70 & 0.52 & $-$0.31 & 0.08 & $-$0.25  \\
    $\Lambda\Sigma^0$ &  0.0 &  0.50 & $-$0.44 & 0.28 & $-$0.04 & 0.29 \\
    \hline
  \end{tabular}

\bigskip 

Table 2: Various fourth order contributions to the magnetic 
         moments (in $\mu_N$). (j) refers to diagram (1h) with
         an insertion $\sim b_{D,F}$. 


\end{center}

\smallskip

\renewcommand{\arraystretch}{1.1}


\begin{center}

\begin{tabular}{|r|r||r|r|r|r|r|r|r|}
    \hline
    $Z$ & ${\cal C}$  & $b_6^D$ & $b_6^F$ &$\alpha_1$ & $\alpha_2$ 
       & $\alpha_3$ &  $\alpha_4$ & $\beta_1$  \\
    \hline
    0   & 1.5 & 3.93 & 3.01 & $-$0.45 & $-$0.06 & $-$0.47 &
              0.45 & $-$0.74 \\    
    0   & 1.2 & 3.82 & 2.95 & $-$0.39 & $-$0.11 & $-$0.45 &
              0.39 & $-$0.67 \\    
    $-$0.3 & 1.5 & 3.83 & 2.95 & $-$0.40 & $-$0.11 & $-$0.45 &
              0.40 & $-$0.67 \\    
    $-$0.3 & 1.2 & 3.76 & 2.92 & $-$0.36 & $-$0.15 & $-$0.43 &
              0.36 & $-$0.62 \\    
         \hline
  \end{tabular}

\bigskip 

Table 3: LECs for various decuplet parameters.


\end{center}

\smallskip

\newpage

\noindent {\Large {\bf Figures}}

$\,$

\vskip 1.5cm

\begin{figure}[h]
\centerline{
\epsfysize=5.2in
\epsffile{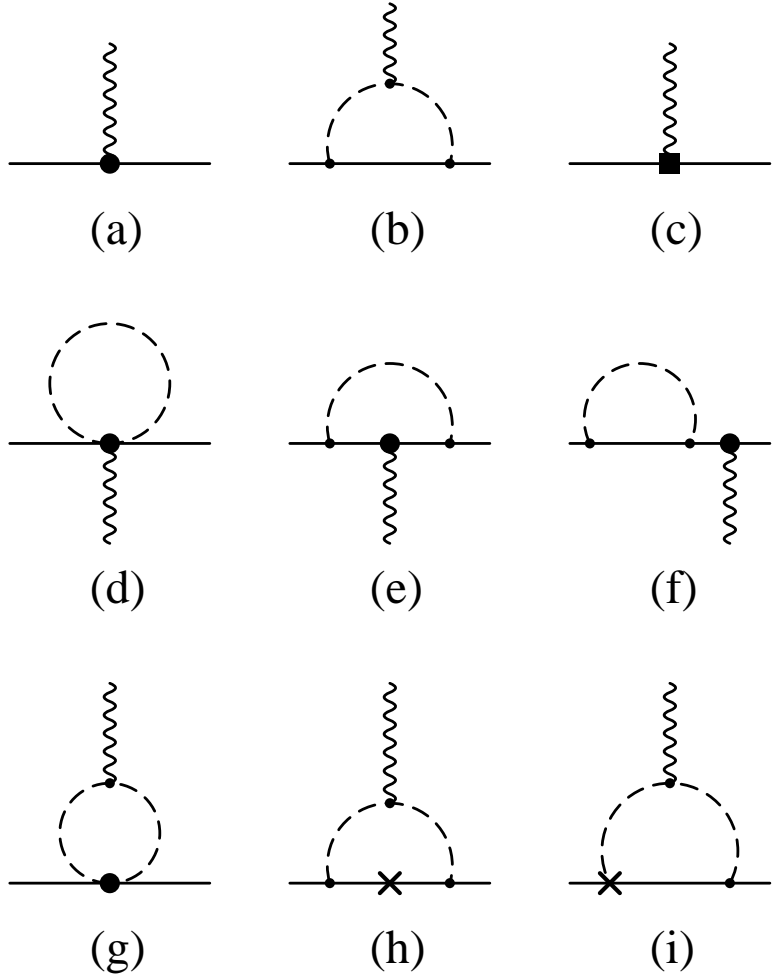}
}
\vskip 0.5cm
\caption{Contribution to the magnetic moments as explained in the
  text. The heavy dots refer to insertions from ${\cal L}^{(2)}_{MB}$
  proportional to the LECs $b_6^D$, $b_6^F$, $b_{9,10,11}$ and the crosses
  refer to an insertion with a fixed coefficient. In case of (h), the
  cross also means an insertion $\sim b_{D,F}$. The box denotes an
  insertion from ${\cal L}^{(4)}_{MB}$. For the wave function
  renormalization (f), only one representative graph is shown. Also,
  the crossed partner to (i) is not depicted.}

\end{figure}

$\,$

\vskip 1cm

\begin{figure}[bht]
\centerline{
\epsfxsize=6in
\epsffile{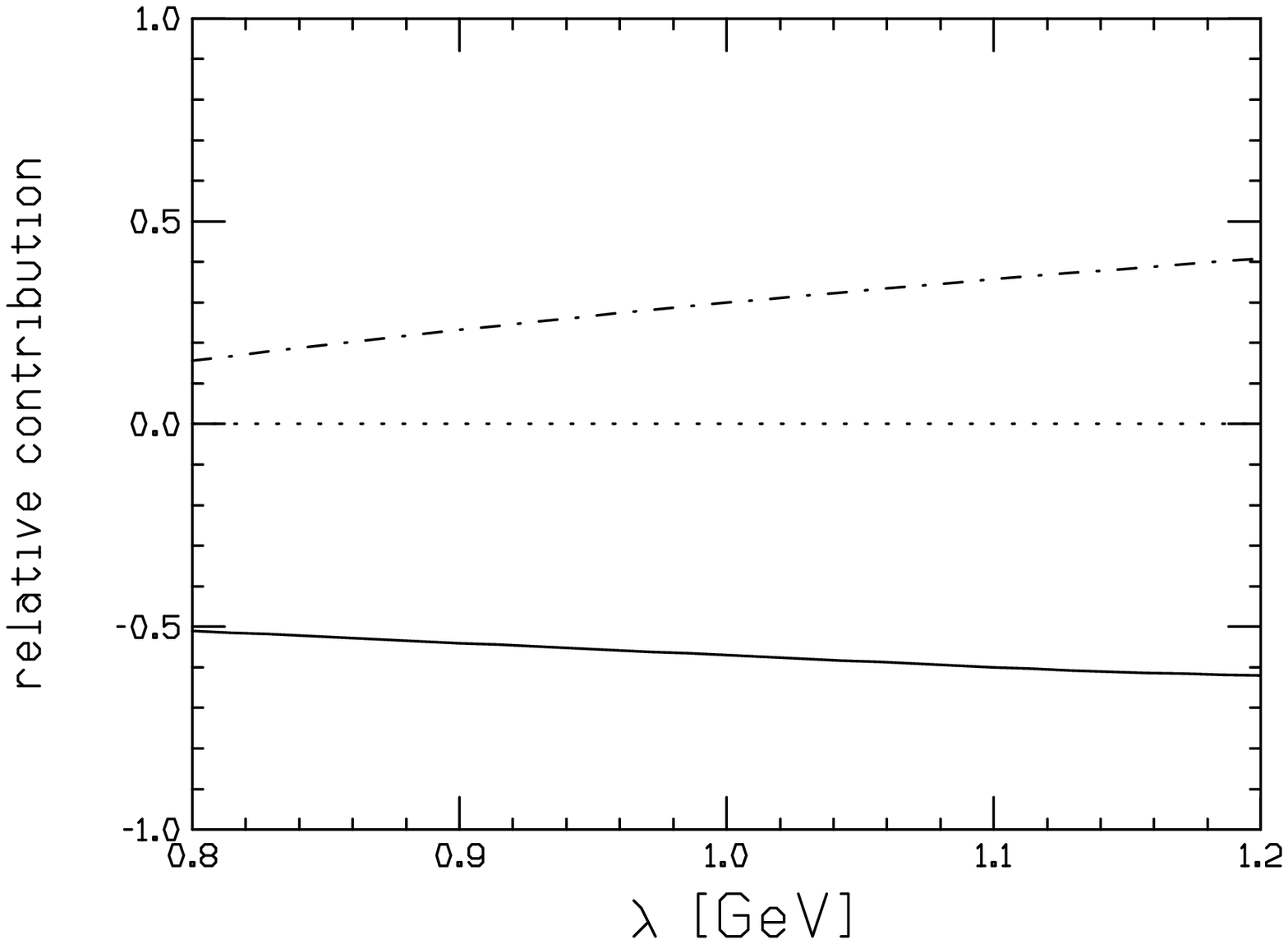}
}
\vskip 1cm
\caption{Relative contributions to the protons magnetic moment 
  as explained in the text. The solid and the dashed--dotted lines 
  refer to the normalized third and fourth order contributions, in order.}
\vskip 1.5cm

\end{figure}


\begin{thebibliography}{99}

\bibitem{CG} S. Coleman and S.L. Glashow, Phys. Rev. Lett. 6 (1961) 423

\bibitem{CP} D.G. Caldi and H. Pagels, Phys. Rev. D10 (1974) 3739

\bibitem{GSS} J. Gasser, M. Sainio and A. ${\rm \check S}$varc, 
Nucl. Phys. B307 (1988) 779

\bibitem{krause} A. Krause, Helv. Phys. Acta 63 (1990) 3

\bibitem{jlms} E. Jenkins, M. Luke, A.V. Manohar and M. Savage, 
Phys. Lett. B302 (1993) 482; 

(E) ibid B388 (1996) 866 

\bibitem{jmNc} E. Jenkins and A.V. Manohar, Phys. Lett. B335 (1994) 452

\bibitem{lmr} M.A. Luty, J. March-Russell and M. White, Phys. Rev. D51
(1995) 2332

\bibitem{bos} J.W. Bos, D. Chang, S.C. Lee, Y.C. Lin and H.H. Shih, preprint
hep-ph/9602251, to be published in Chin. J. Phys. (Taipei) (1997).


\bibitem{jm} E. Jenkins and A.V. Manohar, Phys. Lett. B255 (1991) 558

\bibitem{bkkm}
V. Bernard, N. Kaiser, J. Kambor and Ulf-G. Mei\ss ner, Nucl. Phys.
B388 (1992) 315

\bibitem{bkmz} V. Bernard, N. Kaiser and Ulf-G. Mei\ss ner, Z. Phys.
C60 (1993) 111

\bibitem{bora} B. Borasoy and Ulf-G. Mei{\ss}ner, Ann. Phys. (N.Y.) 
254 (1997) 192

\bibitem{guido} G. M\"uller and Ulf-G. Mei{\ss}ner, 
Nucl. Phys. B492 (1997) 379 

\bibitem{bclls} J.W. Bos, D. Chang, S.C. Lee, Y.C. Lin and H.H. Shih, preprint
hep-ph/9611260

\bibitem{bkml2} V. Bernard, N. Kaiser and Ulf-G. Mei\ss ner, 
Nucl. Phys. A615 (1997) 483

\bibitem{luwi} M.A. Luty and M. White, Phys. Lett. B319 (1994) 261

\bibitem{armin} A. Schmidt, thesis, TU M\"unchen, 1994 (unpublished)

\bibitem{reso} G. Ecker, J. Gasser, A. Pich and E. de Rafael, Nucl. Phys.
B321 (1989) 311

\bibitem{bora2} B. Borasoy and Ulf-G. Mei{\ss}ner, Int. J. Mod. Phys.
A11  (1996) 5138

\bibitem{thom} H. Thom, Phys. Rev. 151 (1966) 1322

\bibitem{liz} E. Jenkins, Nucl. Phys. B368 (1992) 190

\bibitem{lwd} D.B. Leinweber, R.M. Woloshyn and T. Draper,
 Phys. Rev. D43 (1991) 1659 


\end{thebibliography}
\end{document}